\documentclass[%
 reprint,
 amsmath,amssymb,
 aps,
 prd,
]{revtex4-2}

\usepackage{appendix}
\usepackage{graphicx}
\usepackage{dcolumn}
\usepackage{bm}

\usepackage{amssymb}
\usepackage{lipsum}
\usepackage{soul}
\usepackage{slashed}
\usepackage{amsmath}
\usepackage{bbold}
\usepackage[compat=1.0.0]{tikz-feynman}
\usetikzlibrary{decorations.markings,decorations.pathreplacing}

\tikzfeynmanset{
	fermion2/.style={
		/tikz/postaction={
			/tikz/decorate=false,
		},
	},
}
\tikzfeynmanset{
	fermion3/.style={
		/tikz/decoration={
			markings,
			mark=at position 0.5 with {
				\node[
				dot,
				fill,
				draw=none,
				] { };
			},
		},
		/tikz/postaction={
			/tikz/decorate=true,
		},
	},
}
\def\Tr{\textrm{Tr}}

\newcommand{\PiDiagramMom}{%
\raisebox{-0.40\height}{%
\begin{tikzpicture}[scale=0.62, line join=round, line cap=round]
  \tikzset{
    fermionloop/.style={thick, postaction={decorate},
      decoration={markings, mark=at position 0.20 with {\arrow{>}}}},
    pion/.style={densely dotted, thick},
    vertex/.style={circle, fill=black, inner sep=1.0pt}
  }
  \def\r{1.10}
  \def\L{2.35} 

  \draw[pion] (-\L,0) -- (-\r,0) node[midway, above] {$\scriptstyle q$};
  \draw[pion] ( \r,0) -- ( \L,0) node[midway, above] {$\scriptstyle q$};

  \node[vertex] at (-\r,0) {};
  \node[vertex] at ( \r,0) {};

  \draw[fermionloop] (0,0) circle (\r);

  \node at (0,\r+0.45) {$\scriptstyle k-l$};
  \node at (0,-\r-0.45) {$\scriptstyle k+l$};
\end{tikzpicture}}%
}

\DeclareMathOperator*{\SumInt}{%
	\mathchoice%
	{\ooalign{\raisebox{.15\height}{\scalebox{0.9}{$\textstyle\sum$}}\cr\hidewidth$\displaystyle\int$\hidewidth\cr}}
	{\ooalign{\raisebox{.14\height}{\scalebox{.7}{$\textstyle\sum$}}\cr\hidewidth$\textstyle\int$\hidewidth\cr}}
	{\ooalign{\raisebox{.2\height}{\scalebox{.6}{$\scriptstyle\sum$}}\cr$\scriptstyle\int$\cr}}
	{\ooalign{\raisebox{.2\height}{\scalebox{.6}{$\scriptstyle\sum$}}\cr$\scriptstyle\int$\cr}}
}

\begin{document}

\title{Hyperon-Induced Inhomogeneous Pion Condensation \\ and Moat Regimes in Neutron Star Cores}

\author{Theo F. Motta}
\email{theo.motta@unesp.br}
\author{Randall H. V. Pradinett}
\email{randall.vargas@unesp.br}
\author{Gastão Krein}
\email{gastao.krein@unesp.br}
\affiliation{Instituto de Física Teórica, Universidade Estadual Paulista, Rua Dr. Bento Teobaldo Ferraz, 271 - Bloco II - 01140-070 São Paulo, SP, Brazil}

\date{\today}

\begin{abstract}
We perform a stability analysis of the homogeneous ground state of nuclear matter against inhomogeneous perturbations of the pion condensate. In $\beta$-equilibrium, restricting the baryon species to nucleons only, we observe no instability; however, at high densities, the pseudoscalar density-density correlations assume a moat regime, i.e. a damped oscillatory patterned spatial correlation, which in momentum space appears as a non-zero global minimum for some finite three-momentum. When hyperons are permitted to appear, this minimum can cross down to negative values, which configures an instability towards an inhomogeneous pion condensate which ultimately will affect the equation of state.
\end{abstract}

\maketitle

\section{Introduction}
\label{introduction}
Inside the inner-most core of compact stars lie the answers to several open questions in modern nuclear physics. Unfortunately, these objects are not only astronomically far away, they are also unusually small, which makes studying them observationally quite a challenging task. In our favor is the fact that some neutron star properties are so extreme compared with any other physical system that, despite the aforementioned difficulties, it is possible to infer quite a lot about their structure via gravitational and electromagnetic signals. Nevertheless, in order to interpret such measurements, it is of the utmost importance to understand precisely the composition and properties of the dense matter in its interior. 

Among the issues that puzzle the nuclear physics community is whether or not strange matter, either in the form of hyperons (baryons with non-zero strangeness) or deconfined strange quark matter, exists in the core of high mass neutron stars (NS). Naïvely, when one adds such species of particles to the calculation, one finds that most models predict the maximum mass of neutron stars is somewhere between $1.5$M$_\odot$ and $1.7$M$_\odot$ when in fact there have been observations of stars as heavy as $2$M$_\odot$ \cite{Burgio:2021vgk}. Nevertheless, theoretically, strangeness should be energetically favored. This apparent incompatibility of observation with theory is often referred to as the ``hyperon puzzle''. Models that take into account how the internal structure of the baryon is modified by the medium \cite{Motta:2022nlj} often give rise to emergent many-body repulsions that suppress hyperon production at lower to moderate densities, which allows such models to reproduce observed masses, considering the full baryon octet \cite{Motta:2019tjc,Rikovska-Stone:2006gml,Guichon:2018uew}. Broadly speaking, though, the issue of strangeness in stellar matter is, in fact, full of puzzles, all of which currently have several different proposed solutions \cite{Tolos:2020aln}.

Whether or not this or that particle is produced in dense matter is an issue
at the level of the expectation value of single-particle density operators.  Generically, an operator of the type $ \bar\psi_b\mathcal{\hat O}\psi_b$, where $\mathcal{\hat{O}}$ is a matrix in Dirac and isospin space, can be denoted a density operator for the particle $b$, e.g. the occupation number density $\langle \bar\psi_b \gamma_0 \psi_b\rangle$, scalar density $\langle \bar\psi_b\psi_b \rangle$, etc. Lately, the nuclear physics community has started to pay closer attention also to density-density correlation functions, e.g. the static two-point correlation function of the scalar density $\langle \bar\psi(x)\psi(x)\bar\psi(y)\psi(y)\rangle$, occupation number density $\langle \bar\psi(x)\gamma_0\psi(x)\bar\psi(y)\gamma_0\psi(y)\rangle$, etc. They encode important information about the structure of the phase that is not present in the mean-field expectation values of density operators. For instance, three typical behaviors of such correlation functions, (i) trivial and (ii) inhomogeneous, are sketched below:
\begin{center}
    \includegraphics[width=0.45\linewidth]{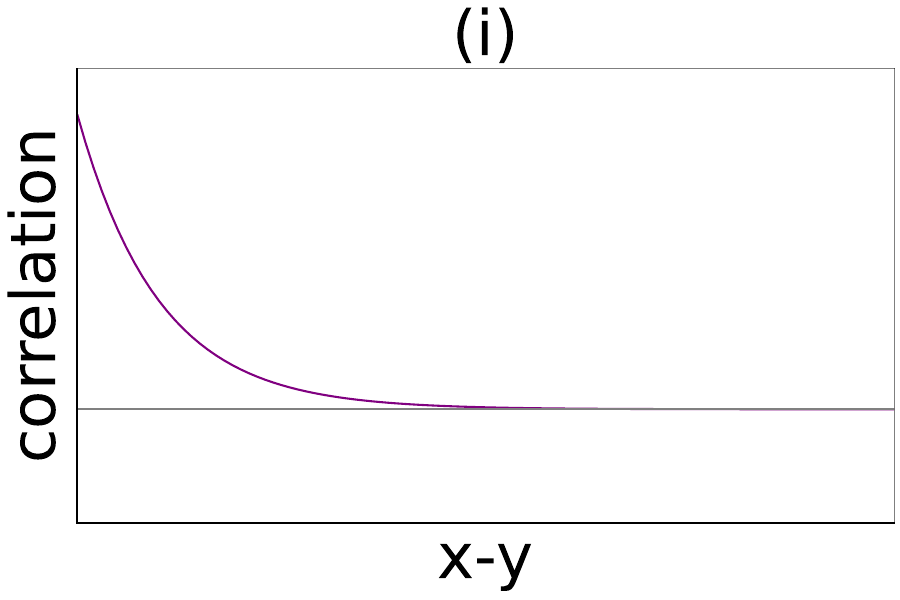}
    \includegraphics[width=0.45\linewidth]{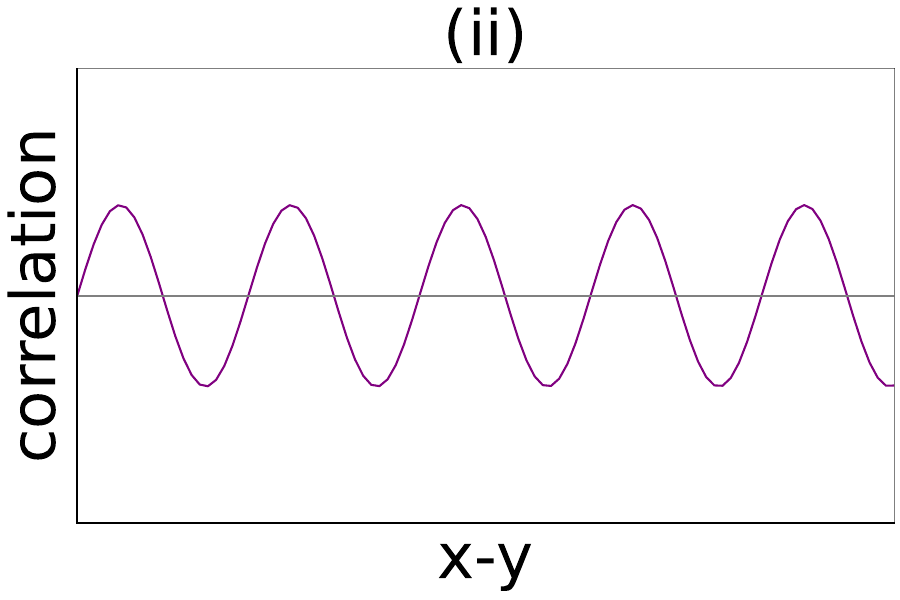}
\end{center}
A ``trivial'' phase will show some monotonic behavior, in the case of a regular Yukawa potential, decay like
\begin{equation}
    \propto \frac{e^{-m|\vec x - \vec y|}}{|\vec x-\vec y|}.
\end{equation}
Inhomogeneous or crystalline phases (ii) will show an oscillatory function with infinitely long range, and this oscillatory behavior would also appear in the one-point function of the relevant local density operator, e.g. $\langle \bar\psi(x)\mathcal{\hat O}\psi(x)\rangle$ will also be oscillatory. These are extremely interesting and, in fact, have been intensely studied theoretically in the past couple of decades \cite{Buballa:2014tba,Buballa:2018hux,Buballa:2020nsi,Buballa:2020xaa,Carignano:2012sx,Carignano:2017meb,Carignano:2018hvn,Motta:2024rvk,Motta:2024agi}. Usually, the focus is on inhomogeneous phases of quark matter that may also be realized in the interior of neutron stars. This is because these phases are brought by an inhomogeneous breaking of chiral symmetry (i.e. the relevant ``density'' operator is the quark condensate). Although they are phenomenologically very rich, in this paper we will deal with stars that do not contain deconfined quark matter in the core. Since the vast majority of neutron stars are not heavier than 1.7M$_\odot$, we shall play it safe and look to what is unambiguously there, which is nuclear matter.

What might be indistinguishable from a trivial phase at the level of the local densities is case number (iii) where the density-density correlations both oscillate and decay. This type of correlation function is characteristic of liquid-crystals and the Quantum-Pion-Liquid (QPL) \cite{Lee:2015bva,Pisarski:2020dnx}, also known as ``complex'' phases, a name which alludes to the fact that, in this phase, the free energy's Hessian matrix eigenvalues \cite{Ogilvie:2024vde,Schindler:2019ugo} appear in complex conjugate pairs. Formally, this is known as a moat regime \cite{Fu_2025,Rennecke:2025kub,cao2025moatregimes21flavor}, since the inverse of this correlation function in momentum space is non-monotonic and shows a depression, referred to as moat, shown in the sketch below
\begin{center}
\includegraphics[width=0.45\linewidth]{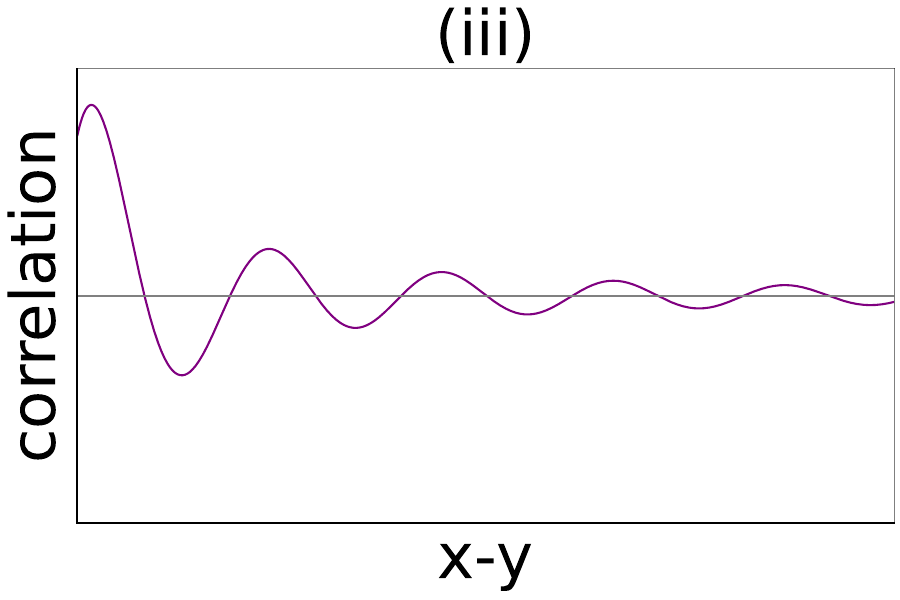}
\includegraphics[width=0.45\linewidth]{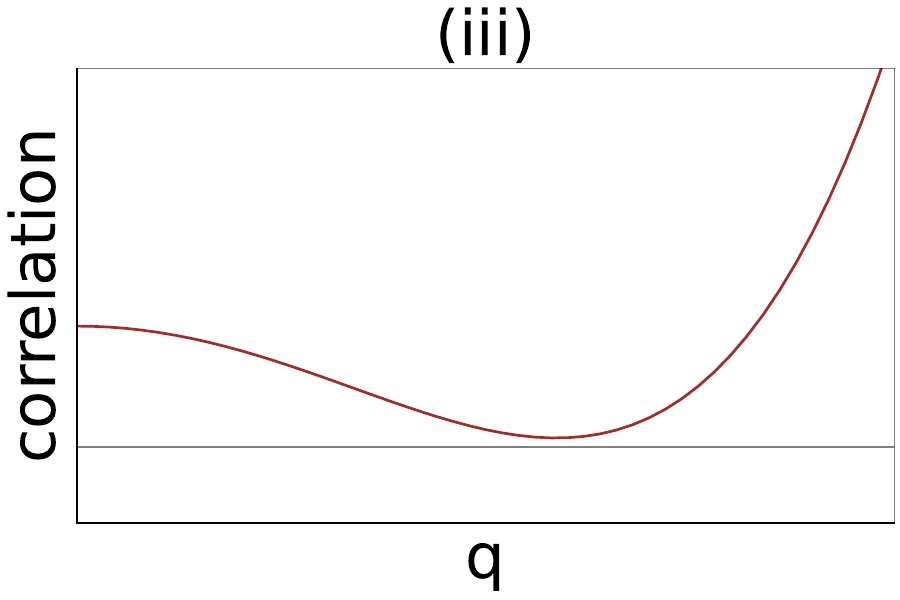}
\end{center}
In the context of heavy-ion collisions, moat regimes have been shown to have experimental signatures that can be looked for in collider experiments \cite{Pisarski:2021qof,Rennecke:2023xhc,Nussinov:2024erh}. Of course, in the context of neutron stars, observations are taken \textit{in natura}, not in controlled experiments. Nevertheless, as stated above, it is important to learn as much as possible about dense matter in order to interpret observations. Especially since gravitational wave astronomy is now a reality, we will hopefully be able to disintangle more and more information about compact star composition. 

There are two main ways to look for inhomogeneous phases. One can either make an Ansatz for the functional form of the expectation values of density operators $f(x)=\langle \bar\psi(x)\mathcal{\hat O}\psi(x)\rangle$ and simply calculate the free energy of the system in this configuration. If the free energy is lower than the homogeneous ground state, then the inhomogeneous state is energetically preferred. Alternatively, one can perform a stability analysis, that is, calculate the homogeneous lowest energy state, introduce a small inhomogeneous perturbation, and calculate what contribution this perturbation brings to the free energy. If it raises the free energy, then the homogeneous ground state is stable, otherwise, if it's negative, it is unstable. As it turns out \cite{Buballa:2018hux,Motta:2025xop} the free energy response to the introduction of this perturbation is intimately tied to the inverse static correlation function: if it becomes negative for \textit{any} finite three-momentum, the spatially trivial ground state is unstable against spatially non-trivial perturbations. In other words, the moat regime configuration can be stable, if the minimum of the inverse correlation function in momentum space is positive. If the minimum crosses the line and becomes negative, the state becomes unstable and an inhomogeneous phase will take its place.

A previous study of \textit{symmetric} nuclear matter \cite{Motta:2025xop} showed that it is unlikely for it to become inhomogeneous or crystalline, at least for {the type of models reviewed in Ref.~\cite{Motta:2022nlj} which consider the modification of the baryon structure in-medium}. In this paper, we extend the calculation to $\beta$-equilibrium matter, considering also the effect of strange baryons in the medium. Studies based on chiral nuclear models \cite{Papadopoulos:2024agt,Papadopoulos:2025uig} have also shown that inhomogeneous phases are disfavored in dense nuclear matter due to charge neutrality and weak equilibrium. Despite their model being somewhat different from the one we use, our calculation confirms the same tendency, but only for the case with only nucleons in the medium. As we will discuss, if hyperons are included, an instability towards inhomogeneous phases appears.

This paper is organized as follows. In Sec.~\ref{Model} we define the microscopic model we use to study baryonic interactions. Following, in Sec.~\ref{Moat}, we discuss in greater depth what the so-called ``moat regime'' is, how we identify it, why it is relevant. In Sec.~\ref{Matter} we briefly discuss neutron star conditions and the inputs behind our calculations. Finally, in Sec.~\ref{Res} we show our results, followed by our discussions and conclusions in Sec.~\ref{Concl}.
 
\section{The Model}
\label{Model}
In the context of neutron star physics, one could define the quark-meson coupling (QMC) model \cite{Guichon:2018uew} by the Lagrangian density
\begin{equation}\label{qmclag}
    \begin{aligned}
        \mathcal{L}&=
    \bar\psi_{b} \big( i\slashed\partial - M_b \big) \psi_{b}
    + g^b_\sigma\bar\psi_{b}\sigma\psi_{b}
    - \frac{g^{b\, 2}_\sigma d^b_\sigma}{2}\bar\psi_{b}\sigma^2\psi_{b}
    \\&- g^b_\omega\bar\psi_{b}\slashed\omega\psi_{b}
    - g^b_\rho\bar\psi_{b}\slashed\rho_a\tau_a\psi_{b}
    - \frac{g_A}{2f_\pi}\bar\psi_{b}\gamma_5\slashed\partial\pi_a\tau_a\psi_{b} 
    \\&+ \mathcal{L}_{\sigma\,\omega\,\rho\,\pi},
    \end{aligned}
\end{equation}
 where $\mathcal{L}_{\sigma\,\omega\,\rho\,\pi}$ accounts for the kinetic mesonic terms, $\psi_b$ represents a baryon (and the index $b$ is implicitly summed over), and $g_\phi^b$ is the baryon's coupling to the mesons $\phi=\sigma,\,\omega,\,\rho$.
Superficially, this differs from traditional Relativistic Mean Field models (RMF) only by the term
\begin{equation}\label{qmcvertex}
\frac{g_\sigma^{b\,2} d^b_\sigma}{2}\bar\psi_{b}\sigma^2\psi_{b}
\end{equation}
which is nothing more than a baryon-antibaryon-sigma-sigma vertex. However, the coupling parameters $g_\sigma^b$ and $d_\sigma$ (the latter is known as the scalar polarizability) are not entirely free parameters. 
In the QMC model, these couplings are obtained from a baryon structure calculation, which is traditionally done by considering the baryon as a bag of three quarks and allowing the quarks to couple to the meson fields directly. Naturally, this induces a non-linear dependence on the baryon properties with respect to the meson mean fields. Most importantly, the coupling of the baryon with the sigma meson becomes dependent on the meson mean field, and one can consider each baryon coupling with the sigma such that via the following vertex
$$
\Gamma_\sigma(\bar\sigma)\bar\psi_{b}\sigma\psi_{b}
$$
where
\begin{equation}\label{vertex}  
    \Gamma_\sigma(\bar\sigma) = g^b_\sigma - \frac{g^{b\,2}_\sigma d^b_\sigma}{2} \bar\sigma.
\end{equation}
The values of the couplings in Eq.~(\ref{vertex}) are extracted from the bag-model calculation. Nevertheless, it is possible to construct the QMC model based on any hadron structure framework. The review article \cite{Motta:2022nlj} compiles several of these QMC-like models and discusses their differences. In this work, we adopt the most conventional formulation. We do not, however, anticipate any qualitative modification of the results were an alternative, yet reasonable, description of hadron structure to be employed.

Ultimately, the model has a total of five parameters, namely, the radius of the bag, the mass of the scalar channel $m_\sigma$, and the couplings of the mesons with the nucleon, which we will write without a $b$ index: $g_\sigma$, $g_\omega$ and $g_\rho$. 
For the current work, we take the expressions for how the baryon masses depend on the sigma mean field from Ref.~\cite{Rikovska-Stone:2006gml} 
\begin{equation}\label{stone}
\begin{aligned}
M_N^\star(\sigma)& =  M_N-g_\sigma \sigma\\&+\left[0.0022+0.1055 R_{\text {b }}-0.0178R_{\text {b}}^2\right]\left(g_\sigma \sigma\right)^2 \\
M^\star_{\Lambda}(\sigma)& =  M_{\Lambda}-\left[0.6672+0.0462 R_{\text {b }}-0.0021R_{\text {b}}^2\right] g_\sigma \sigma \\
& +\left[0.0016+0.0686 R_{\text {b }}-0.0084R_{\text {b}}^2\right]\left(g_\sigma \sigma\right)^2 \\
M_{\Sigma}^\star(\sigma)& =  M_{\Sigma}-\left[0.6706-0.0638 R_{\text {b }}-0.008R_{\text {b}}^2\right] g_\sigma \sigma \\
& +\left[-0.0007+0.0786 R_{\text {b }}-0.0181R_{\text {b}}^2\right]\left(g_\sigma \sigma\right)^2 \\
M_{\Xi}^\star(\sigma)& =  M_{\Xi}-\left[0.3395+0.0282 R_{\text {b }}-0.0128R_{\text {b}}^2\right] g_\sigma \sigma \\
& +\left[-0.0014+0.0416 R_{\text {b }}-0.0061R_{\text {b}}^2\right]\left(g_\sigma \sigma\right)^2,
\end{aligned}
\end{equation}
from which we can extract the sigma coupling parameters for each baryon, since the effective mass in the mean field can be written as
\begin{equation}
    M^\star_b(\sigma) = M_b -g_\sigma^b\sigma + \frac{d_\sigma^b}{2}(g_\sigma^b\sigma)^2.
\end{equation}
Clearly, we can see that the couplings of the mesons to the hyperons, as well as the characteristic scalar polarizability $d_\sigma^b$, are all dependent on the size of the baryon ($R_b$ in Eq.~(\ref{stone})), which we have chosen to set at $0.7$fm. The $\sigma$ mass is fixed at $600$MeV. Finally, we have the other meson masses
\begin{equation}
    m_\omega = 783~\text{MeV},\quad
    m_\rho = 770~\text{MeV},\quad
    m_\pi=140~\text{MeV},
\end{equation}
the pion decay constant $f_\pi=93$~MeV, the axial-vector coupling constant $g_A=1.26$, 
and the three remaining parameters are the couplings $g_{\sigma, \,\omega,\, \rho}$ which we will fix to standard nuclear matter properties.

The model turns out to be extremely successful in describing finite nuclei properties \cite{Stone:2016qmi}, and neutron-star properties as well \cite{Motta:2019tjc,Kalaitzis:2019dqc,Leong:2023yma}. In a previous reference, we showed that it also washes away the instabilities of the homogeneous ground state of the Walecka model \cite{Serot:1992ti}. For more information on the model, readers should look for references \cite{Motta:2022nlj,Leong:2023yma,Motta:2019tjc,Guichon:1987jp,Guichon:1995ue} that define and explain the model in greater detail. For our current intent, it should suffice to say that we take Eq.~(\ref{qmclag}) to define our theory of meson-mediated baryon-baryon interactions, together with the couplings taken from Eq.~(\ref{stone}).

\section{The Moat}
\label{Moat}
As alluded to in the introduction, the moat regime can be defined directly by the behavior of the inverse static correlation function in momentum space, and we can calculate it via the zero-frequency limit of the inverse meson propagator of the relevant collective mode.

In the relativistic description of dense nuclear matter, each meson is dressed by particle-hole excitations of the baryonic medium.
The full inverse meson propagator in Random-Phase-Approximation (RPA) is defined via the Dyson equation which can be written diagramatically as
\begin{equation}\label{eq:dse}
    \begin{tikzpicture}
        \begin{feynman}
            \vertex (a);
            \vertex [right=0.5cm of a] (ap);
            \vertex [right=1cm of a] (b);
            \diagram*{
                (a) -- [boson] (b);
            };
            \draw (ap) node [dot];
        \end{feynman}
    \end{tikzpicture}^{-1}
    \,\,=\,\,
    \begin{tikzpicture}
        \begin{feynman}
            \vertex (a);
            \vertex [right=0.5cm of a] (ap);
            \vertex [right=1cm of a] (b);
            \diagram*{
                (a) -- [boson] (b);
            };
        \end{feynman}
    \end{tikzpicture}^{-1}
    \,\,-\,\,
    \raisebox{-0.27cm}{
    \begin{tikzpicture}
    \begin{feynman}
    \vertex (a);
    \vertex [right=0.25cm of a] (ap);
    \vertex [right=0.5cm of ap] (bp);
    \vertex [right=0.25cm of bp] (b);
    \vertex [above= 0.2cm of ap] (up);
    \vertex [below= 0.2cm of ap] (down);
    \vertex [right=0.25cm of up] (t1);
    \vertex [right=0.25cm of down] (t2);
    \diagram*{
    	(a) --  [boson] (ap);
    	(ap) -- [fermion2, half left] (bp);
    	(ap) -- [fermion2, half right] (bp);
    	(bp) -- [boson] (b);
    };
    \draw (t1) node [dot];
    \draw (t2) node [dot];
    \end{feynman}
    \end{tikzpicture}}\,,
\end{equation}
where the dotted and undotted wiggly lines represent dressed and undressed meson propagators and the second term, the self-energy, contains dressed baryon propagators shown as dotted solid lines.
This inverse correlation function will be denoted $D^{-1}_\phi(q_0,\Vec{q})$ for the modes $\phi=\pi,\,\sigma,\,\omega,\,\rho$. Whenever the zero-frequency limit of $D^{-1}_\phi$ is not monotonic, this configures a moat regime.

\subsection{Inhomogeneous Instabilities}
Within the scope of the moat-regime, the most drastic scenario happens when the inverse meson propagators have a negative global minimum for some finite three-momentum. This indicates that the homogeneous ground state is unstable with respect to the creation of inhomogeneous condensates \cite{Motta:2025xop,Buballa:2014tba}. However, since there is mixing in the scalar and vector channels at non-zero densities, it is best to write the isoscalar modes as one matrix-valued mesonic channel, such that in free space
\begin{equation}
    \begin{tikzpicture}
        \begin{feynman}
            \vertex (a);
            \vertex [right=0.5cm of a] (ap);
            \vertex [right=1cm of a] (b);
            \diagram*{
                (a) -- [boson] (b);
            };
        \end{feynman}
    \end{tikzpicture}=
    \left(\begin{matrix}
        \Delta^0 & 0 \\
        0 & D^0_{\mu\nu}
    \end{matrix}\right),
\end{equation}
where
\begin{equation}
    \begin{aligned}
    \Delta^0(q)&=\frac{1}{q^2-m_\sigma^2},
    \\
    D^0(q)&=\frac{-1}{q^2-m_\omega^2},
    \\
    D^0_{\mu\nu}(q) &= {\left(\eta_{\mu\nu}-\frac{q_\mu q_\nu}{q^2}\right)}
    D^0(q),
    \end{aligned}
\end{equation}
and the {self-energy contains} off-diagonal terms that correspond to the mixings 
\begin{equation}\label{mixings}
    \raisebox{-0.25cm}{
    \begin{tikzpicture}
    \begin{feynman}
    \vertex (a);
    \vertex [right=0.25cm of a] (ap);
    \vertex [right=0.5cm of ap] (bp);
    \vertex [right=0.25cm of bp] (b);
    \vertex [above= 0.2cm of ap] (up);
    \vertex [below= 0.2cm of ap] (down);
    \vertex [right=0.25cm of up] (t1);
    \vertex [right=0.25cm of down] (t2);
    \diagram*{
    	(a) --  [boson] (ap);
    	(ap) -- [fermion2, half left] (bp);
    	(ap) -- [fermion2, half right] (bp);
    	(bp) -- [boson] (b);
    };
    \draw (t1) node [dot];
    \draw (t2) node [dot];
    \end{feynman}
    \end{tikzpicture}
    }=
    \left(\begin{matrix}
        \Pi_s & \Pi_{\mu\phantom{\mu}} \\
        \Pi_{\nu} & \Pi_{\mu\nu}
    \end{matrix}\right).
\end{equation}
As we discussed in \cite{Motta:2025xop}, in this scenario where mixing is involved, one option to look for inhomogeneous instabilities of any of the modes together is to make use of the Static Dielectric Function (SDF), defined by
\begin{equation}
    \epsilon(\vec{q})=
    \text{det}\left(
    \mathbb{1}-
    \raisebox{-0.22cm}{
    \begin{tikzpicture}
    \begin{feynman}
    \vertex (a);
    \vertex [right=0.90cm of a] (ap);
    \vertex [right=0.5cm of ap] (bp);
    \vertex [right=0.25cm of bp] (b);
    \vertex [above= 0.2cm of ap] (up);
    \vertex [below= 0.2cm of ap] (down);
    \vertex [right=0.25cm of up] (t1);
    \vertex [right=0.25cm of down] (t2);
    \diagram*{
    	(a) --  [boson] (ap);
    	(ap) -- [fermion2, half left] (bp);
    	(ap) -- [fermion2, half right] (bp);
    	(bp) -- [boson] (b);
    };
    \draw (t1) node [dot];
    \draw (t2) node [dot];
    \end{feynman}
    \end{tikzpicture}}
    \right)_{q_0\rightarrow0^+}.
\end{equation}
If this function is non-monotonic, one or more of the eigenvalues of the inverse propagator matrix must also be non-monotonic, and this configures a moat regime. Most importantly, though, zeros of the SDF imply instability of the homogeneous ground state with respect to small inhomogeneous fluctuations oscillating periodically with wave vector $\vec q$ \cite{Buballa:2014tba,Nickel:2009ke,Nickel:2009wj,Broniowski:2011ef,Deryagin:1992rw}.

In this setup, the $\pi$ and the $\rho$ do not mix with the scalar sector or with each other. Therefore, for the isovector case, it is sufficient to simply look at their static inverse propagator. Ultimately, whenever
$$\epsilon(\Vec{q}_\text{min})<0,\quad\text{or}\quad
D_{\pi,\rho}(q_0=0,\Vec{q}_\text{min})<0$$
for some finite $\Vec{q}$, this region is unstable and that the true ground state is inhomogeneous.

\section{The Matter}
\label{Matter}

Our goal is to model neutron star matter. Therefore, take Eq.~(\ref{qmclag}) and calculate its equation of state. 
This is traditionally done in mean-field, dressing the baryons by
\begin{equation}\label{dse}
    \raisebox{-0.0cm}{
    \begin{tikzpicture}
    \begin{feynman}
    \vertex (a);
    \vertex [left=0.50cm of a] (b);
    \vertex [right=0.50cm of a] (c);
    \diagram*{
    	(c) --  [fermion2] (b);
    };
    \draw (a) node [dot];
    \end{feynman}
    \end{tikzpicture}}^{-1}
    =
    \raisebox{-0.0cm}{
    \begin{tikzpicture}
    \begin{feynman}
    \vertex (a);
    \vertex [left=0.50cm of a] (b);
    \vertex [right=0.50cm of a] (c);
    \diagram*{
    	(c) --  [fermion2] (b);
    };
    \end{feynman}
    \end{tikzpicture}}^{-1}
    +
    \raisebox{-0.0cm}{
    \begin{tikzpicture}
    \begin{feynman}
    \vertex (a);
    \vertex [left=0.50cm of a] (b);
    \vertex [right=0.50cm of a] (c);
    \vertex [above=0.50cm of a] (d);
    \vertex [above=0.40cm of d] (d1);
    \diagram*{
    	(c) --  [fermion2] (b);
        (a) --  [boson] (d);
        (d) --  [fermion2, half left] (d1);
        (d) --  [fermion2, half right] (d1);
    };
    \draw (d1) node [dot];
    \end{feynman}
    \end{tikzpicture}}\,,
\end{equation}
where the in-medium baryon propagators are
\begin{equation}
    \label{fermprop2}
    \begin{aligned}
        G(k) =& \Bigg[
    \frac{\slashed k^\star + M^\star}{k^{\star\,2} - M^{\star\,2} +i\epsilon}
    +\\&
    \frac{\slashed k^\star + M^\star}{E^\star(\vec k)}i\pi \theta(k_F - |\vec{k}|)\delta(k^\star_0-E^\star(\vec k))
    \Bigg].
    \end{aligned}
\end{equation}
and their masses are dressed by
\begin{equation}
    M^\star(\bar\sigma)_b=
    M_b -\Gamma^b_\sigma(\bar\sigma)\bar\sigma,
\end{equation}
where $\Gamma^b_\sigma$ is defined by Eq.~(\ref{vertex}).
In that sense, the baryon propagator in a medium separates into a vacuum part and a density-dependent part, 
\begin{equation}\label{f+d}
    G(k) = G_F(k) + G_D(k),
\end{equation}
where $G_F$ is the usual Feynman propagator (virtual particle-antiparticle fluctuations), and $G_D$ encodes the filled Fermi sea (real particles with $\vec k \leq k_F$).

Note that, after inserting this decomposition into each component of the meson polarizations Eq.~(\ref{mixings}), they separate into four physically distinct contributions
\begin{equation}
    \Pi = \Pi_{\rm FF} + \Pi_{\rm FD} + \Pi_{\rm DF} + \Pi_{\rm DD}
\end{equation}
coming from the contributions of Feynman (F) and Density dependent (D) propagators.
The $\rm FD$ and $\rm DF$ terms describe a process where a baryon from the Fermi sea is excited into an empty state above the Fermi surface, which is called particle-hole excitations. 
The $\rm DD$ term corresponds to the fluctuation of the medium constituents inside the Fermi sea. However, in the static limit ($q_0 \to 0$), this contribution is entirely suppressed in the mesonic polarization.

For any mesonic modes $a,b$ with couplings $g_a, g_b$ 
and Dirac vertices $\Gamma_a, \Gamma_b$, the \textit{medium--dependent} part of the 
polarization can be written as
\begin{equation}\label{PI}
\begin{aligned}
\Pi_{ab}(q)
= -i g_a g_b 
\int \frac{d^{4}k}{(2\pi)^{4}}\,
&\mathrm{tr}\Big[
\,\Gamma_a S_{F}\!\left(k+\tfrac{q}{2}\right)\Gamma_b S_{D}\!\left(k-\tfrac{q}{2}\right)
\\[0.2em]
&+\,\Gamma_a S_{D}\!\left(k+\tfrac{q}{2}\right)\Gamma_b S_{F}\!\left(k-\tfrac{q}{2}\right)
\\[0.2em]
&+\,\Gamma_a S_{D}\!\left(k+\tfrac{q}{2}\right)\Gamma_b S_{D}\!\left(k-\tfrac{q}{2}\right)
\Big].
\end{aligned}
\end{equation}
The FF term, which accounts for vacuum fluctuations, is already present at zero density. However, vacuum fluctuations in hadronic models are usually discarded, and doing so configures the so-called ``no-sea'' approximation. Since this is a low energy effective model of baryons, it makes sense to discard the $\Pi_{\rm FF}$ retaining only the density-dependent pieces $\Pi_{\rm FD}$, $\Pi_{\rm DF}$, $\Pi_{\rm DD}$, which are responsible for all in-medium effects studied in this paper. A more detailed discussion on Eq.~(\ref{PI}) and its explicit calculation is shown in Appendix~\ref{apdx} for the most consequential channel, the pion.

We then calculate the energy density $\mathcal{E}$ including baryons exchange terms, which a function of the density of each fermion species $\{n_b\}=n_{N^0}\,,n_{N^+ }\,,n_\Lambda\,
,n_{\Sigma^0}\,,n_{\Sigma^-}\,,n_{\Sigma^+}\,
,n_{\Xi^0}\,,n_{\Xi^-}$, i.e.
\begin{equation}
    \begin{aligned}
        \mathcal{E}(\{n_b\})=
    \SumInt_{\vec k} \sqrt{k^2+M_b^{\star 2}}+
    \raisebox{-0.22cm}{
    \begin{tikzpicture}
    \begin{feynman}
    \vertex (a);
    \vertex [right=0.60cm of a] (ap);
    \vertex [right=0.5cm of ap] (bp);
    \vertex [left=0.5cm of a] (cp);
    \diagram*{
    	(a) --  [boson] (ap);
    	(ap) -- [fermion2, half left] (bp);
    	(ap) -- [fermion2, half right] (bp);
    	(cp) -- [fermion2, half left] (a);
    	(cp) -- [fermion2, half right] (a);
    };
    \draw (bp) node [dot];
    \draw (cp) node [dot];
    \end{feynman}
    \end{tikzpicture}}
    +
    \raisebox{-0.4cm}{
    \begin{tikzpicture}
    \begin{feynman}
    \vertex (a);
    \vertex [right=0.40cm of a] (xp);
    \vertex [above=0.350cm of xp] (up);
    \vertex [below=0.350cm of xp] (dp);
    \vertex [right=0.80cm of a] (ap);
    \diagram*{
    	(a) --  [boson] (ap);
    	(a) -- [fermion2, half left] (ap);
    	(a) -- [fermion2, half right] (ap);
    };
    \draw (up) node [dot];
    \draw (dp) node [dot];
    \end{feynman}
    \end{tikzpicture}},
    \end{aligned}
\end{equation}
where the last diagram corresponds to a Fock term. Note that the corresponding Fock diagram to the baryon equation (Eq.~(\ref{dse})) is absent, however, this is a reasonable approximation since its effect is much smaller to the baryon then to the energy density and it would raise the numerical price of the calculation substantially.
Finally, having defined our energy density, the next goal becomes finding what densities of each species minimize the free energy under the constraints of $\beta$-equilibrium and charge neutrality, for a fixed total baryon number $n_B$. 

\section{Results}
\label{Res}

\begin{figure}
    \centering
    \includegraphics[width=\linewidth]{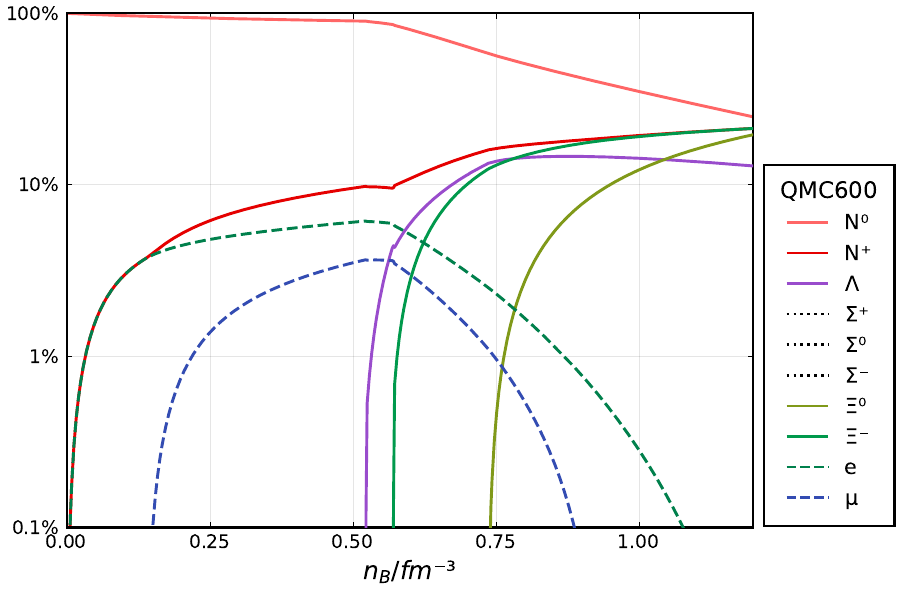}
    
    \includegraphics[width=\linewidth]{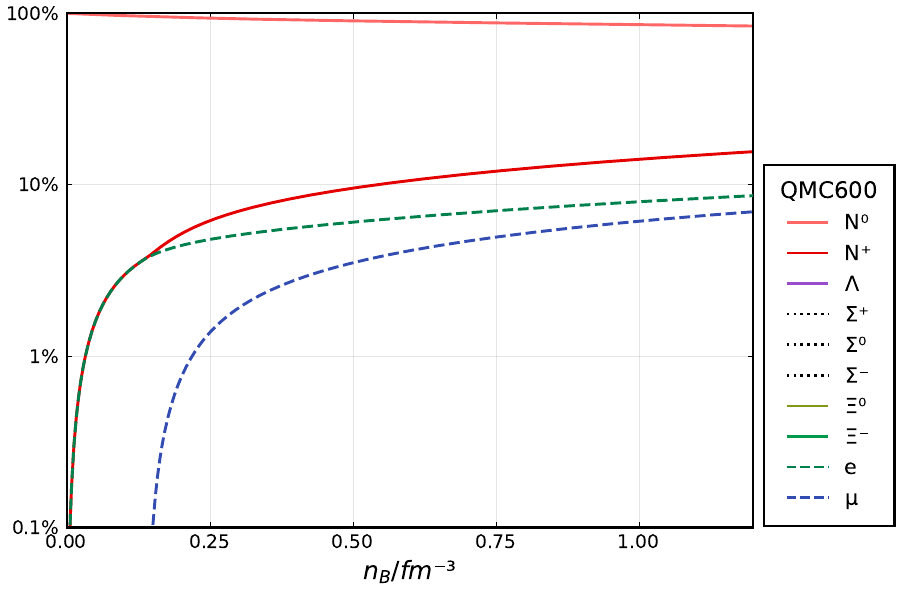}
    \caption{Densities of each fermion species shown in proportion to the total baryon number. Both plots taken with $m_\sigma = 600$~MeV. On the left we see the result including the full baryon octet and, on the right, we show the species fraction for nucleons only (plus electrons and muons).}
    \label{fig:densities}
\end{figure}

The couplings are fixed to reproduce standard nuclear matter properties at saturation, i.e.
\begin{equation}
    n_\text{sat}=0.16\text{fm}^{-3},\quad
    E/A = -16\text{MeV},\quad a_\text{sym}=30\text{MeV}.
\end{equation}
In table~\ref{tab:params} we show the coupling constants for two different values of the sigma mass. Since the sigma channel is truly an effective channel (accounting for $f_0$ exchange, correlated and uncorrelated two-pion exchanges, etc), we show couplings for two different $\sigma$ masses. However, the effect of this choice on our results is minimal, as we will discuss shortly. Figure~\ref{fig:densities} shows density of each baryon and lepton species in proportion to the total baryon density. In the upper plot, the results include the full baryon octet, and in the lower plot nucleons only. Note that it is at about 0.5~fm$^{-3}$ that the hyperons kick in, first the $\Lambda$ and then the cascades. This fact will be relevant for our discussion further on.
\begin{table}
    \centering
    \begin{tabular}{|c|c|c|c|c|}\hline
        QMC & $g_\sigma$ & $g_\omega$  & $g_\rho$ 
        \\ \hline 
       \,$ m_\sigma \phantom{\Big|}$= 600 MeV\,  & \, 9.98 \, & \, 10.45 \, & \, 6.92\,
        \\
        $ m_\sigma $ = 700 MeV & \, 11.67 \, & \, 10.35\, & \,6.47\,
        \\
        \hline
    \end{tabular}
    \caption{The mass of the $\omega$ channel is kept at $783$~MeV and the mass of the $\rho$ is kept at $770$~MeV}
    \label{tab:params}
\end{table}
Now, by taking the $\beta$-equilibrium densities, shown in Fig.~\ref{fig:densities}, we can ultimately calculate both the isoscalar SDF and the inverse static propagators of the isovector channels as a function of the total baryon number density.

\begin{figure}
    \centering
    \includegraphics[width=\linewidth]{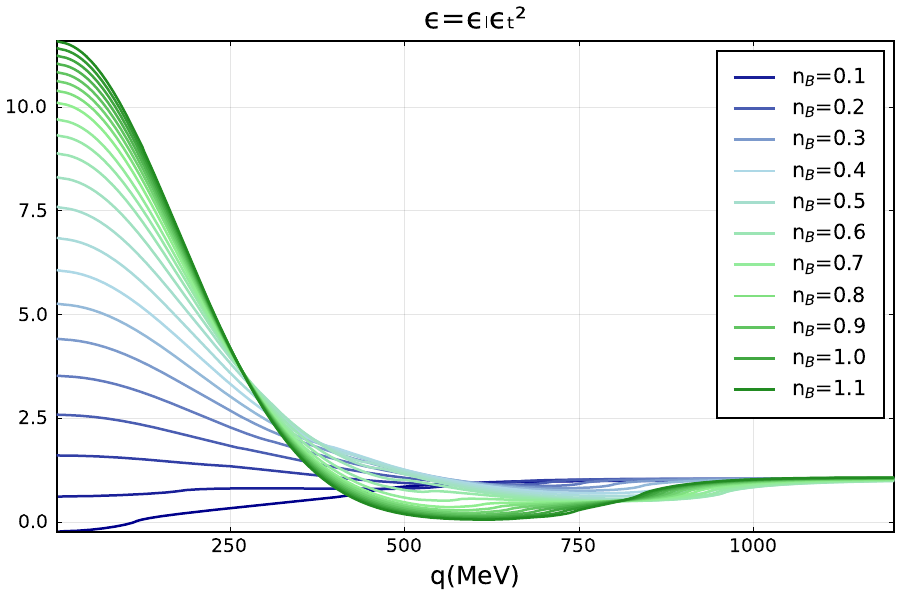}
    \caption{Isoscalar SDF, including the full baryon octet, in $\beta$-equilibrium.}
    \label{fig:sdf}
\end{figure}

For instance, in Fig.~\ref{fig:pions0} we show the $\pi^0$ two-point function for the case without the hyperons, for several different densities. This means that for every total baryon density in Fig.~\ref{fig:pions0}, the individual species densities are taken from Fig.~\ref{fig:densities} for the case with nucleons only. Clearly, Fig.~\ref{fig:pions0} shows a vast region in density where the static propagator is non monotonic. This starts to happen shortly before the density hits $n_\text{B}\approx 0.5$fm$^{-3}$. Nevertheless, the moat minima never get below zero, indicating that although a moat regime is in place, there is no instability towards an inhomogeneous phase. With the full baryon octet, however, Fig.~\ref{fig:pions1} shows that in $\beta$-equilibrium, we do see a large region where the static two-point function goes below zero. This means that an instability of the homogeneous ground state towards the generation of inhomogeneous $\pi^0$ condensation is present. Naturally, in such a case, there is little point in observing any other mesonic channel since, according to our calculation, the true ground state is crystalline. In other words, the fact that a $\pi^0$ condensate breaks translational symmetry spontaneously will affect other meson condensates, likely turning them inhomogeneous as well. This negativity, however, happens only after $n_\text{B}\approx 0.6$fm$^{-3}$. For lower densities, the system could still be in a moat-regime, and this is in fact what we find, especially for the isoscalar sector. The SDF of the isoscalar, shown in Fig.~\ref{fig:sdf}, sector indeed has a global non-zero minimum for densities as low as $0.15$fm$^{-3}$. 
It becomes non-monotonic very quickly and even at saturation density, a moat-regime is in place.

\section{Discussion and Conclusions}
\label{Concl}

Our results impel  us to conclude that the matter in the core of a neutron star of moderate mass, up to M=1.7~M$_\odot$, {resides} in a moat regime. Nuclear matter in $\beta$ equilibrium assumes non-trivial spatial density-density correlations in the isoscalar sector as early as $0.15$~fm$^{-3}$. It remains to be seen, however, whether this will have observational consequences as the moat regime does in heavy-ion collisions \cite{Pisarski:2021qof}. If it does have consequences, this could open up a new avenue in neutron star phenomenology and it warrants further research. 
\begin{figure}
    \centering
    \includegraphics[width=\linewidth]{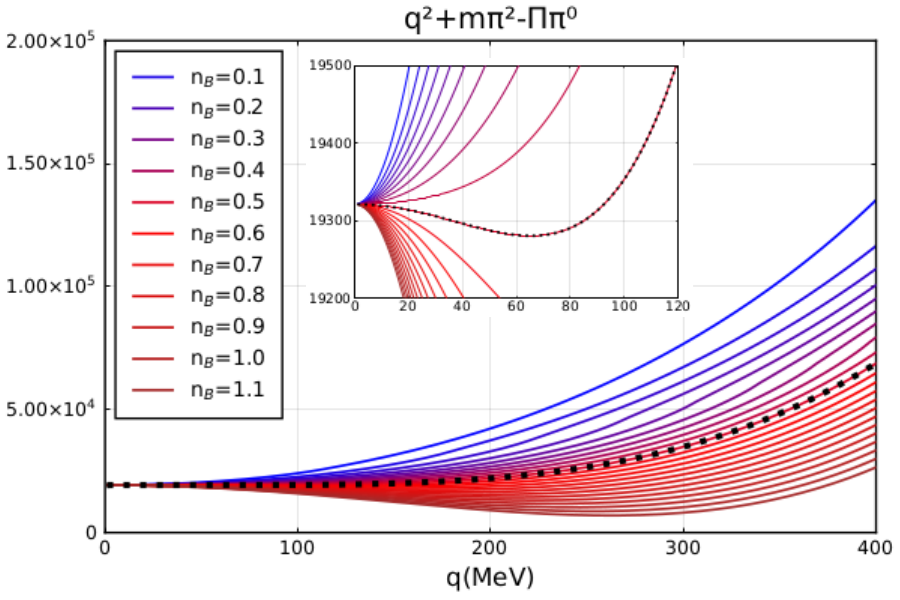}
    \caption{Inverse pion two-point functions for different densities without hyperons. The dotted line shows the first line with a negative inclination at the origin, i.e., the first density to manifest a moat regime $n_B\approx$0.5~fm$^{-3}$.}\label{fig:pions0}
\end{figure}
\begin{figure}
    \centering
    \includegraphics[width=\linewidth]{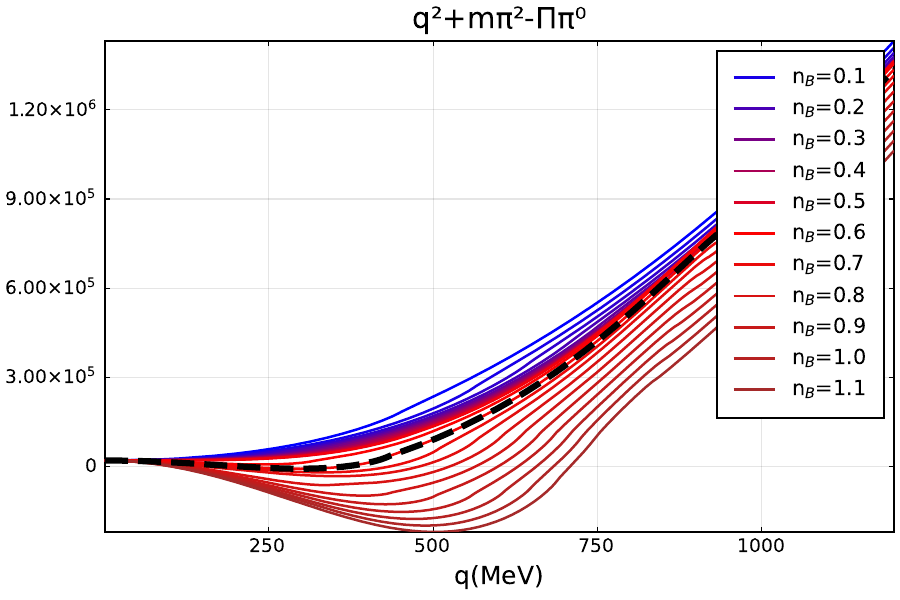}
    \caption{Same as Fig.~\ref{fig:pions0}, now including the full baryon octet. Up till a density of $n_B\approx$0.5~fm$^{-3}$, which is where the moat sets in, matter consists of nucleons only. After that there are deviations from Fig.~\ref{fig:pions0} and eventually, at a density of $n_B\approx$0.65~fm$^{-3}$, shown in the dashed line, the minima begin to cross zero.}
    \label{fig:pions1}
\end{figure}

By far, the most interesting result is the fact that the isovector-pseudovector density-density correlations, i.e., pion correlations, become negative for large densities, which indicates the presence of inhomogeneous pion condensation. Moreover, and strikingly, this \textit{only} happens in the presence of hyperons. Figs \ref{fig:pions0} and \ref{fig:pions1} naturally show the same curves for densities below and up to $0.5$~fm$^{-3}$ since, before this point, $\beta$-equilibrium nuclear matter consists only of nucleons. After this point, the fate of the system drastically changes depending on whether we allow hyperons to appear or not. In the baryon-octet complete case, this develops a deep moat regime which eventually reaches zero, defining an instability. With nucleons only, this moat regime is not deep enough. Note that at this point, all we are saying is that the homogeneous ground state is unstable with respect to crystallization. The equation of state of nuclear matter will certainly change \cite{Buballa:2014tba} and affect neutron star structure.

Naturally, whether or not an inhomogeneous phase is stable will be model-dependent. However, the fact that it \textit{could} be present in neutron star cores is what is important. In Ref.~\cite{Papadopoulos:2024agt}, for instance, nuclear matter in a chiral density wave phase was investigated and shown to be disfavored in a chiral nuclear model. Since inhomogeneous phases might be more available observationally from gravitational and electromagnetic probes of neutron stars (at a minimum because the equation of state is affected, but it certainly could affect other properties as well), this could be used to discriminate between models. If inhomogeneous phases \textit{are} found to be present in neutron star cores, models that predict them to be unstable or disfavored are invalidated.

\section*{Acknowledgements}
We would like to thank our colleague Fabian Rennecke for enlightening discussions.
This work was partially financed by Funda\c{c}\~ao de Amparo \`a Pesquisa do Estado de S\~ao Paulo (FAPESP), grant nos. 2024/13426-0 (TFM) and 2018/25225-9 (GK), and Conselho Nacional de Desenvolvimento Cient{\'i}fico e Tecnol{\'o}gico (CNPq), grant nos. 315225/2025-4 (TFM), 88887.893649/2023-00 (RHVP) and 309262/2019-4 (GK).

\appendix 
\section{Pion polarization}\label{apdx}
We start from the standard pseudovector pion nucleon interaction Lagrangian,
with all momenta taken incoming,
\begin{equation}
\mathcal{L}_{\pi N}(x)=
-\frac{g_A}{2f_\pi}\,
\bar\psi(x)\,
\gamma_5\gamma^\mu
\left(\partial_\mu \pi^a(x)\right)
\tau^a\,
\psi(x),
\end{equation}
where $\psi=(p,n)^T$ is the nucleon isodoublet, $\pi^a$ ($a=1,2,3$) is the pion isotriplet, and $\tau^a$ are the Pauli matrices in isospin space.

Fourier transforming the pion field,
$\partial_\mu\pi^a(x)\to -iq_\mu \pi^a(q)$,
and removing the overall momentum-conserving delta function from Dyson's expansion,
the amputated momentum-space vertex reads:
\begin{equation}
{
i\Gamma^a(q)=
-i\,\frac{g_A}{2f_\pi}\,
\gamma_5\slashed{q}\,
\tau^a,
}
\qquad
(q^\mu \text{ incoming at the vertex}).
\end{equation}

At zero temperature and finite density, the nucleon propagator for each species
$i=p,n$ can be decomposed into a vacuum part and a density-dependent contribution,
\begin{equation}
S_i(k)=S_F(k)+S_D^{(i)}(k).
\end{equation}
Using effective mean-field quantities $k^{*\mu}$ and $M^*$, the vacuum propagator is
\begin{equation}
S_F(k)=
\frac{\slashed{k}^*+M^*}{k^{*2}-M^{*2}+i\epsilon},
\end{equation}
while the density-dependent part places the nucleon on shell and restricts its momentum to the Fermi sea,
\begin{equation}
\begin{aligned}
    S_D^{(i)}(k)&=
i\pi\,
\frac{\slashed{k}^*+M^*}{E^*(\mathbf{k})}\,
\theta\!\big(k_F^{(i)}-|\mathbf{k}|\big)\,
\delta\!\big(k_0^*-E^*(\mathbf{k})\big),
\\
E^*(\mathbf{k})&=\sqrt{\mathbf{k}^2+M^{*2}}.
\end{aligned}
\end{equation}
The dependence on the nuclear composition enters solely through the Fermi momenta
$k_F^{(p)}$ and $k_F^{(n)}$.

The pion polarization in a given isospin channel $a$ is given by the one-loop diagram with two pseudovector vertices,
\begin{equation}
    \PiDiagramMom
\end{equation}
i.e.
\begin{equation}
\begin{aligned}
    -i\left(\frac{g_A}{2f_\pi}\right)^2
\sum_{i,j=p,n}
\int\frac{d^4k}{(2\pi)^4}\,
(\tau^a_{ij}\tau^a_{ji})\,
\Tr_D\!\Big[
&S_i(k-l)\,
\gamma_5\slashed{q}
\\&\times S_j(k+l)
\gamma_5\slashed{q}
\Big].
\end{aligned}
\end{equation}
where we introduced the symmetric momentum routing
\begin{equation}
l^\mu\equiv\frac{q^\mu}{2},
\qquad
\ell\equiv|\boldsymbol{\ell}|=\frac{|\mathbf{q}|}{2},
\qquad
l^2=l_0^2-\boldsymbol{\ell}^{\,2}.
\end{equation}

The isospin factor $\tau^a_{ij}\tau^a_{ji}$ determines which nucleon species contribute:
\begin{itemize}
\item $\pi^0$: $(i,j)=(p,p)$ and $(n,n)$;
\item $\pi^+$: $(i,j)=(n,p)$;
\item $\pi^-$: $(i,j)=(p,n)$.
\end{itemize}

The Dirac trace appearing in the loop can be written in axial form as
\begin{equation}
T_A^{\mu\nu}(k,l)=
\Tr_D\!\left[
(\slashed{k}^*-\slashed{l}+M^*)
\gamma_5\gamma^\mu
(\slashed{k}^*+\slashed{l}+M^*)
\gamma_5\gamma^\nu
\right].
\end{equation}
Using $\gamma_5\gamma^\mu\gamma_5=-\gamma^\mu$ and contracting with $q_\mu q_\nu$,
with $q=2l$, the numerator can be written in compact form as
\begin{equation}
\begin{aligned}
    N_\pi(k,l)&\equiv
q_\mu q_\nu T_A^{\mu\nu}(k,l)
\\&=
16\Big[
2(l\!\cdot\!k^*)^2
-l^4
-l^2 k^{*2}
-l^2 M^{*2}
\Big].
\end{aligned}
\end{equation}
This expression is common to all medium-dependent contributions.

Now, we can expand the propagators as $S_i=S_F+S_D^{(i)}$ leads to
\begin{equation}
\Pi=\Pi_{FF}+\Pi_{FD}+\Pi_{DF}+\Pi_{DD}.
\end{equation}
For medium functions, only the density-dependent terms
$\Pi_{FD}$, $\Pi_{DF}$ and $\Pi_{DD}$ are relevant.
The purely vacuum contribution $\Pi_{FF}$ is omitted, since it does not encode medium effects and would require renormalization.

The $FD$ term corresponds to the left propagator in vacuum and the right one on shell,
\[
\Pi^{(j)}_{FD}:
\quad
S_i(k-l)\to S_F,
\qquad
S_j(k+l)\to S_D^{(j)}.
\]
Using the delta function in $S_D^{(j)}$ to perform the $k_0$ integration,
and shifting $\mathbf{v}=\mathbf{k}+\boldsymbol{\ell}$,
the angular integration can be carried out analytically.
Defining $E(v)=\sqrt{v^2+M^{*2}}$, one obtains
\begin{equation}
\begin{aligned}
    \Pi^{(j)}_{FD}(q)&=
\frac{g_A^2}{(2\pi)^2 f_\pi^2}
\int_0^{k_F^{(j)}} dv\,
\frac{v^2}{E(v)}
\bigg[
C_0^{FD}(v)
\\&+
C_1^{FD}(v)
\ln\!\left|
\frac{l_0^2-l_0E(v)-\ell^2+v\ell}
     {l_0^2-l_0E(v)-\ell^2-v\ell}
\right|
\bigg],
\end{aligned}
\end{equation}
with
\begin{equation}
{
C_0^{FD}(v)=-8l_0,
\qquad
C_1^{FD}(v)=
-\frac{4\,l^2 M^{*2}}{E(v)\,v\,\ell}.
}
\end{equation}

The $DF$ term is obtained analogously by interchanging the roles of the two propagators,
\[
\Pi^{(i)}_{DF}:
\quad
S_i(k-l)\to S_D^{(i)},
\qquad
S_j(k+l)\to S_F,
\]
and performing the shift $\mathbf{v}=\mathbf{k}-\boldsymbol{\ell}$.
The result is
\begin{equation}
\begin{aligned}
    \Pi^{(i)}_{DF}(q)&=
    \frac{g_A^2}{(2\pi)^2 f_\pi^2}
    \int_0^{k_F^{(i)}} dv\,
    \frac{v^2}{E(v)}
    \bigg[
    C_0^{DF}(v)
    \\
    &+
    C_1^{DF}(v)
    \ln\!\left|
    \frac{l_0^2+l_0E(v)-\ell^2+v\ell}
         {l_0^2+l_0E(v)-\ell^2-v\ell}
    \right|
    \bigg],
\end{aligned}
\end{equation}
with
\begin{equation}
{
C_0^{DF}(v)=+8l_0,
\qquad
C_1^{DF}(v)=
-\frac{4\,l^2 M^{*2}}{E(v)\,v\,\ell}.
}
\end{equation}

These expressions are numerically efficient because all angular dependence is contained in a single logarithm, leaving only a one-dimensional integral over the Fermi sea.

When both propagators are density dependent, the polarization is controlled by step functions enforcing Pauli blocking,
\begin{equation}
\begin{aligned}
    \Pi_{DD}^{(ij)}(q)=
-2i\,\frac{g_A^2}{f_\pi^2}\,
\frac{M^{*2}l^2}{\ell}
\int_0^{k_F^{(i)}}& dv\,
\frac{v}{E(v)}\,
\\&\times\Theta\!\left(
E_F^{*(j)}-E(v)-2l_0
\right)\\&\times
\left[
1-\left|
\frac{l_0E(v)+l_0^2-\ell^2}{v\ell}
\right|
\right],
\end{aligned}
\end{equation}
where $E_F^{*(j)}=\sqrt{(k_F^{(j)})^2+M^{*2}}$.
Is important to mention that for the static limit, this contribution vanishes and is included only for completeness.

Finally, the polarization for each physical pion is obtained by fixing the loop indices according to isospin,
\begin{equation}
\begin{aligned}
    \Pi_{\pi^0}(q)&=\Pi_{pp}(q)+\Pi_{nn}(q),
\\
\Pi_{\pi^+}(q)&=\Pi_{np}(q),
\\
\Pi_{\pi^-}(q)&=\Pi_{pn}(q).
\end{aligned}
\end{equation}
Within each $(i,j)$ channel,
\begin{equation}
\Pi_{ij}^{\mathrm{med}}(q)=
\Pi_{FD}^{(j)}(q)+
\Pi_{DF}^{(i)}(q)+
\Pi_{DD}^{(ij)}(q).
\end{equation}

\nocite{ancfiles}

\bibliographystyle{ieeetr} 
\bibliography{pioninstability}

\end{document}